\documentclass[10pt]{JHEP3}
\usepackage{epsfig,multicol,bbm}
\usepackage{epsfig}
\usepackage{amsmath}
\usepackage{epsfig}
\usepackage{psfrag}
\usepackage{amsfonts}
\usepackage{graphicx}
\usepackage{dcolumn}
\usepackage{bm}
\usepackage{lineno}
\usepackage{amssymb}
\usepackage{txfonts}
\usepackage{epsfig}
\usepackage{psfrag}
\usepackage{amsfonts}
\usepackage{graphicx}
\usepackage{dcolumn}
\usepackage{bm}

\newcommand{\fft}[2]{{\frac{#1}{#2}}}

\title{
The Subleading Term of the Strong Coupling Expansion of the Heavy-Quark Potential
in a $\mathcal N=4$ Super Yang-Mills Plasma}
\author{Zi-qiang Zhang $^{\dagger a}$, De-fu Hou$^{\dagger a}$,
 Hai-cang Ren $^{\dagger b}$,$^{\dagger a}$  and Lei Yin$^{\dagger a}$
 \\
{$^{\dagger a}$ Institute of Particle Physics, Huazhong Normal
University, Wuhan 430079, China}\\
{E-mail:zhangzq@iopp.ccnu.edu.cn, ~hdf@iopp.ccnu.edu.cn, ~lei@iopp.ccnu.edu.cn }\\
{$^{\dagger b}$Physics Department, The Rockefeller University,
1230 York Avenue, New York, NY 10021-6399} \\
{E-mail:~ren@mail.rockefeller.edu}\\
 }

\abstract{Applying the AdS/CFT correspondence, the expansion of the
heavy-quark potential of the ${\cal N}=4$ supersymmetric Yang-Mills
theory at large $N_c$ is carried out to the sub-leading term in the
large 't Hooft coupling at a nonzero temperature. The strong
coupling corresponds to the semi-classical expansion of the
string-sigma model, the gravity dual of the Wilson loop operator,
with the sub-leading term expressed in terms of functional
determinants of fluctuations. The contribution of these determinants
are evaluated numerically.}

\keywords{AdS/CFT  correspondence , Heavy quark potential, Finite temperature}
\begin{document}
\section{Introduction}

A prominent implication of the AdS/CFT duality
\cite{Maldacena:1997re,Gubser:1998bc,Witten:1998qj,MadalcenaReview}
is the correspondence between the type IIB superstring theory
formulated on AdS$_5\times S^5$ and $\mathcal N=4$ supersymmetric
Yang-Mills theory (SYM) in four dimensions. In particular, the
supergravity limit of the string theory corresponds to the leading
behavior of SYM at large $N_c$ and large 't Hooft coupling
\begin{equation}
\lambda \equiv g_{\rm YM}^2N_c = \fft{L^4}{\alpha'^2}.
\label{eq:ldef}
\end{equation}
with $L$ the AdS radius and $\alpha'$ the reciprocal of the string tension. This
relation thereby has been applied fruitfully to explore the strongly coupled QGP created in RHIC in spite of its underlying dynamics, QCD, is different from $\mathcal N=4$ SYM.
See \cite{SonReview} for a review and the references therein. It is
expected some of the properties of the latter is universal for all strongly interacting systems.
The heavy quark potential addressed in this paper is one example of such an expedition.
It is expected that some discoveries of $\mathcal N=4$ SYM at large $N_c$ and large $\lambda$ may shed new light on real QCD and other strongly interacting systems (For a recent comparison with the lattice simulation of the large $N_c$ QCD, see \cite{MarcoReview}).

The heavy quark potential of QCD is an important
quantity that probes the confinement mechanism in the hadronic phase and
the meson melting in the plasma phase. It is extracted from the expectation
of a Wilson loop operator
\begin{equation}
W(C)=Pe^{-i\oint_C dx_\mu A_\mu}
\end{equation}
with $A_\mu$ the gauge potential and the symbol $P$ enforcing the
path ordering along the loop $C$. The thermal expectation value
$<W(C)>$ can be measured for QCD on a lattice and the heavy quark
potential is defined with F-ansatz or U-ansatz. The F-ansatz
accounts for the induced free energy of a heavy quark and a heavy
antiquark separated by a distance $r$
\begin{equation}
F(r,T)=-T\ln\frac{<{\rm tr}W({\cal C}_2)>}{<{\rm tr}W({\cal C}_1)>^2},
\label{fansatz}
\end{equation}
where ${\cal C}_1$ is a straight line running in the direction of
the Euclidean time, closed by the periodicity of the Matsubara
formulation, and ${\cal C}_2$ consists of two such lines separated
by $r$ and running in opposite directions. The self-energy of each
heavy quark has been subtracted via $W({\cal C}_1)$. The U-ansatz
accounts for the induced internal energy and is given by
\begin{equation}
U(r,T)=-T^2\frac{\partial}{\partial T}\left(\frac{F}{T}\right).
\label{uansatz}
\end{equation}
We have $F(r,0)=U(r,0)$.

In the case of $\mathcal N=4$ SYM, the AdS/CFT duality relates the
Wilson loop expectation value to the path integral of the
string-sigma action developed in \cite{Metsaev:1998it} of the
worldsheet in the AdS$_5\times S^5$ bulk spanned by the loop on the
boundary for $T=0$ and its generalization in the presence of a black
hole for $T>0$ \cite{Cvetic:2000}. To the leading order of strong
coupling, the path integral is given by its classical limit, which
is the minimum area of the world sheet, and the heavy quark
potential of F-ansatz extracted for the loop $C_2$ and $C_1$ reads
\cite{Maldacena:1998im} \cite{Rey:1998bq}\cite{Liu2}
\begin{equation}
F(r,T)=-\frac{4\pi^2}{\Gamma^4\left(\frac{1}{4}\right)}\frac{\sqrt{\lambda}}{r}
{\rm min}[g_0(rT),0],
\label{leading}
\end{equation}
where $g_0(rT)$ is a monotonic decreasing function and is given by
(\ref{g0}) of the next section. We have $g_0(0)=1$ and $g_0(r_0T)=0$
with the screening length $r_0\simeq \frac{0.7541}{\pi T}$. The
leading order potential of the U-ansatz is obtained by substituting
(\ref{leading}) into (\ref{uansatz}).

The strong coupling expansion of the SYM Wilson loop corresponds to the semi-classical
expansion of the string-sigma action and is parameterized by
\begin{equation}
F(r,T)=-\frac{4\pi^2}{\Gamma^4\left(\frac{1}{4}\right)}\frac{\sqrt{\lambda}}{r}
{\rm min}\Big[g_0(rT)+\frac{\kappa g_1(rT)}{\sqrt{\lambda}}+O\left(\frac{1}{\lambda}\right),0\Big]
\label{expansion}
\end{equation}
with $g_1(0)=1$. The coefficient $\kappa$, corresponding to the
subleading term at $T=0$, has been calculated numerically in
\cite{chu:2009} and analytically in \cite{forini:2010}. We have
$\kappa\simeq -1.33460$. Computing the function $g_1(rT)$ is the
main scope of this work. We find that $g_1(rT)$ is a monotonically
decreasing function, which reaches 0.92 at $r_0$ where
$g_0(r_0T)=0$. Therefore the temperature correction to the
subleading term of the heavy quark potential is numerically small.
The screening radius to the subleading order, $r_0^\prime$,
determined by
\begin{equation}
g_0(r_0^\prime T)+\frac{\kappa g_1(r_0^\prime T)}{\sqrt{\lambda}}=0
\end{equation}
reads
\begin{equation}
r_0^\prime=r_0-\frac{0.267267}{\sqrt{\lambda}T}.
\end{equation}

This work is the continuation of the previous calculation at zero
temperature \cite{chu:2009}. The classical solution of the
string-sigma model and the one loop effective action underlying
$\kappa$ is briefly reviewed in the next section. There we also
outline our strategy of computation, which is within the framework
of \cite{chu:2009} and \cite{Hou:2009zk}. The computation of the
functional determinants involved by Gelfand-Yaglom method
\cite{Gelfand} is described in the sections III and IV. The section
V concludes the paper along with some discussions of the result and
some open issues.


\section{The formulation}


Let us begin with a brief review of the classical limit that leads to the leading order
potential (\ref{leading}). The string-sigma action in this limit reduces to the
Nambu-Goto action
\begin{equation}
S_{\rm NG}=\frac{1}{2\pi\alpha^\prime}\int d^2\sigma\sqrt{g},
\label{NG}
\end{equation}
with $g$ the determinant of the induced metric on the string world sheet embedded in the target space, i.e.
\begin{equation}
g_{\alpha\beta}=G_{\mu\nu}\frac{\partial X^\mu}{\partial\sigma^\alpha}
\frac{\partial X^\nu}{\partial\sigma^\beta}
\end{equation}
where $X^\mu$ and $G_{\mu\nu}$ are the target space coordinates and
the metric, and $\sigma^\alpha$ with $\alpha=0,1$ parameterize the
world sheet. The target space here is Schwarzschild-AdS$_5\times
S^5$, whose metric may be written as
\begin{equation}
ds^2=\frac{1}{z^2}(fdt^2+d\vec x^2+\frac{dz^2}{f})+d\Omega_5^2,
\label{target}
\end{equation}
where $f=1-\frac{z^4}{z_h^4}$ with $z_h=\frac{1}{\pi T}$, and
$d\Omega_5$ is the element of the solid angle of S$^5$. The physical
3-brane resides on the AdS boundary $z=0$. The string world sheets
considered in this paper are all projected onto a point of S$^5$ in
the classical limit.

The Wilson loop of a static heavy quark, denoted by ${\cal C}_1$, is
a straight line winding up the Euclidean time $0<t<\beta$
periodically at the AdS boundary, where $\beta=1/T$. The
corresponding world sheet in the AdS bulk can be parameterized by
$t$ and $z$ with $\vec x$ constant and extends all the way to
Schwarzschild horizon, $z\to{z_h}$. The induced metric is given by
\begin{equation}
ds^2[{\cal C}_1]=\frac{1}{z^2}(fdt^2+\frac{1}{f}dz^2) \label{1quark}
\end{equation}
with the scalar curvature in a diagonal metric
\begin{equation}
R=-\frac{1}{\sqrt{g}}\frac{d}{dz}{\frac{\frac{dg_{00}}{dz}}{\sqrt{g}}}
\label{R}.
\end{equation}
We find
\begin{equation}
R=-2(1-3\frac{z^4}{z_h^4})\equiv R^{(1)}. \label{r1}
\end{equation}
Substituting the metric (\ref{1quark}) into (\ref{NG}), we find the self-energy of
the heavy quark
\begin{equation}
E[{\cal C}_1]=TS^c[{\cal
C}_1]=\frac{1}{2\pi\alpha^\prime}\int_\varepsilon^{z_h}\frac{dz}{z^2}.
\label{s1}
\end{equation}
Notice that we have pulled the physical brane slightly off the
boundary with the radial coordinate $z=\varepsilon$, as a
regularization of the divergence pertaining the lower limit of the
integral (\ref{s1}).

The free energy of a pair of a heavy quark and a heavy antiquark
separated by a distance $r$ can be extracted from the Wilson loop
consisting of two anti-parallel lines each winding up the Euclidean
time at the boundary. This Wilson loop will be denoted by ${\cal
C}_2$ and the world sheet in the bulk can be parameterized by $t$
and $z$ with $x^1, x^2={\rm const.}$ The function $x^3(z)$ is
determined by substituting the induced metric
\begin{equation}
ds^2[{\cal
C}_2]=\frac{1}{z^2}\{fdt^2+\Big[\left(\frac{dx^3}{dz}\right)^2+1\Big]dz^2\},
\label{s2}
\end{equation}
into the action (\ref{NG}) and minimizing it. We have
\begin{equation}
x^3=\pm\int_z^{z_0}dz'\frac{z'^2\sqrt{z_h^4-z_0^4}}{\sqrt{(z_0^4-z'^4)(z_h^4-z'^4)}}.
\label{xi}
\end{equation}
The maximum bulk extension of the world sheet, $z_0<z_h$, is determined by the distance $r$
between the two lines at the boundary and we find that
\begin{equation}
r=\frac{2\sqrt{z_h^4-z_0^4}}{z_0}\int_0^1\frac{\zeta^2d\zeta}{\sqrt{(1-\zeta^4)(s^4-\zeta^4)}}
\label{z0}
\end{equation}
where $s=\frac{z_h}{z_0}$. Substituting (\ref{xi}) into ({\ref{s2}),
we end up with the induced metric
\begin{equation}
ds^2[{\cal
C}_2]=\frac{1}{z^2}\left(fdt^2+\frac{z_0^4}{z_0^4-z^4}dz^2\right),
\label{ss2}
\end{equation}
and the scalar curvature
\begin{equation}
R=-2+2\frac{1-f}{1-f_0}(3-\frac{4f_0}{f^2})\equiv R^{(2)} \label{r2}
\end{equation}
where $f_0=1-\frac{z_0^4}{z_h^4}$. The free energy of the heavy
quark pair is given by
\begin{equation}
E[{\cal C}_2]=TS^c[{\cal
C}_2]=\frac{1}{\pi\alpha^\prime}z_0^2\int_\varepsilon^{z_0}\frac{\sqrt{f}dz}{z^2\sqrt{z_0^4-z^4}},
\label{e2}
\end{equation}
where the same regularization is applied to the lower limit of the integral.

The function $g_0(rT)$ of (\ref{leading}) that underlies the heavy
quark potential is obtained by subtracting from (\ref{e2}) the self
energy of each quark(antiquark), i.e.
\begin{equation}
-\frac{4\pi^2}{\Gamma^4\left(\frac{1}{4}\right)}\frac{\sqrt{\lambda}}{r}
g_0(rT)\equiv\lim_{\varepsilon\to 0^+}(E[{\cal C}_2]-2E[{\cal
C}_1])=\frac{1}{\pi\alpha^\prime}\Big[\int_0^{z_0}dz
\left(\frac{\sqrt{f}z_0^2}{z^2\sqrt{z_0^4-z^4}}-\frac{1}{z^2}\right)
-\int_{z_0}^{z_h}\frac{dz}{z^2}\Big], \label{g0}
\end{equation}
and is divergence free.

The one loop effective action, $W$ is obtained by expanding the
string-sigma action of \cite{Cvetic:2000} to the quadratic order of
the fluctuating coordinates around the minimum area and carrying out
the path integral \cite{Hou:2009zk}. We have
\begin{equation}
W[{\cal
C}_1]=-\ln\Big[\frac{{\det}^4(-i\gamma^\alpha\nabla_\alpha+\tau_3)}
{{\det}^{\frac{3}{2}}(-\nabla^2+\frac{8}{3}+R^{(1)}){\det}^{\frac{5}{2}}(-\nabla^2)}
\Big], \label{straight}
\end{equation}
for the static quark or antiquark and
\begin{equation}
W[{\cal C}_2]=-\ln\Big[\frac{{\rm
det}^4(-i\gamma^\alpha\nabla_\alpha+\tau_3)}
{{\det}^{\frac{1}{2}}(-\nabla^2+4+R^{(2)}-2\delta)\,
{\det}(-\nabla^2+2+\delta)\,{\det}^{\frac{5}{2}}(-\nabla^2)}\Big], \label{parallel}
\end{equation}
for the quark pair, where
\begin{equation}
\delta=\frac{2{z_h}^4}{z_0^4}(\frac{1}{f}-1)(f-f_0)
\end{equation}
The determinants in the denominators of (\ref{straight}) and
(\ref{parallel}) come from the fluctuations of three transverse
coordinates of the Schwarzschild-AdS sector and five coordinates of
$S^5$ with the Laplacian given by the metric (\ref{1quark}) or
(\ref{ss2}). The determinants in the numerators come from the
fermioninc fluctuations, where we have introduced the 2d gamma
matrices, $\gamma_0=\gamma^0=\tau_2$, $\gamma_1=\gamma^1=\tau_1$ and
$\gamma_0\gamma_1=-i\tau_3$ with $\tau_1$, $\tau_2$ and $\tau_3$ the
three Pauli matrices. In terms of the zweibein of the world sheet,
$e_\alpha^j$, we have $\gamma_\alpha\equiv e_\alpha^j\gamma_j$ with
$j=0,1$ and the covariant derivative
\begin{equation}
\nabla_\alpha=\frac{\partial}{\partial x_\alpha}+\frac{1}{8}
[\gamma_i,\gamma_j]\omega_\alpha^{ij}
\end{equation}
with $\omega_\alpha^{ij}$ the spin connection corresponding to
(\ref{1quark}) or (\ref{ss2}). The power "4" comes from eight 2d
Majorana fermions each of which contributes a power 1/2. The one
loop correction to the heavy quark potential is then
\begin{equation}
F_1(r,T)\equiv T\lim_{\varepsilon\to 0^+}(W[{\cal C}_2]-2W[{\cal
C}_1]). \label{oneloop}
\end{equation}

The simplicity of the static gauge (Nambu-Goto action) adopted here
is not cost free. The effective action $W[{\cal C}_1]$ or $W[{\cal
C}_2]$ logarithmically divergent, the coefficient of which is
proportional to the volume part of the Euler character
\begin{equation}
\int_{z>\varepsilon}dt dz\sqrt{g}R \label{euler}
\end{equation}
of each world sheet with the same coefficient of proportionality
\cite{Drukker:2000ep}. It follows from (\ref{1quark}), (\ref{r1}),
(\ref{ss2}) and (\ref{r2}) that the $O(\frac{1}{\epsilon})$ term of
the integral (\ref{euler}) for the parallel lines is exactly twice
of that for the single line. We have indeed that
\begin{equation}
\int
d^2\sigma\sqrt{g}R=\beta\int_\varepsilon^\infty\frac{dz}{z^2}[-2(1-3\frac{z^4}{z_h^4})]
=-\frac{2\beta}{\varepsilon}+\frac{4\beta}{z_h} \label{euler1}
\end{equation}
for the straight line and
\begin{eqnarray}
\int
d^2\sigma\sqrt{g}R &=& 2\beta\int_\varepsilon^{z_0}dz\frac{z_0^2\sqrt{f}}{z^2\sqrt{(z_0^4-z^4})}
[-2+2\frac{1-f}{1-f_0}(3-\frac{4f_0}{f^2})]\nonumber\\
&=&\frac{2\sqrt{z_0^4-z^4}}{z_0^2z\sqrt{f}}(2-f)\mid_\varepsilon^{z_0}=-\frac{4\beta}{\varepsilon}+O(\varepsilon^3).
\label{euler2}
\end{eqnarray}
for the parallel lines. Unlike the zero temperature case, however,
the logarithmic divergence does not cancel in the combination of
(\ref{oneloop}) in the limit $\varepsilon\to0$ because of the 2nd
term on RHS of (\ref{euler1}). This may be the artifact of different
topologies of ${\cal C}_2$ and two ${\cal C}_1$'s in the presence of
the back hole horizon. The integral (\ref{euler2}) does not match
twice of (\ref{euler1}) as $z_0\to z_h$, when the metrics
(\ref{1quark}) and (\ref{ss2}) agree. Nevertheless, a finite
one-loop correction to the heavy quark potential can still be
extracted by the following argument, which is slightly handwaving
\cite{Hou:2009zk}. Assuming that the one-loop formulation of
\cite{Drukker:2000ep} in the Polyakov gauge can be generalized to
the nonzero temperature case and the effective actions computed in
this gauge, denoted by ${\cal W}[{\cal C}_2]$ and ${\cal W}[{\cal
C}_1]$, are both finite and the one-loop correction reads
\begin{equation}
F_1(r,T)\equiv T({\cal W}[{\cal C}_2]-2{\cal W}[{\cal C}_1]).
\label{oneloopP}
\end{equation}
It was shown in \cite{Drukker:2000ep} that ${\cal W}[{\cal C}_1]=0$
at $T=0$. Because $T$ is the only dimensional quantity for ${\cal
W}[{\cal C}_1]$, it must remain zero for all $T$. Next, we introduce
a reference temperature $T_0$ and the corresponding world sheet with
the superscript "(0)", we have
\begin{equation}
F_1(r,T)-F_1(r,T_0)=T{\cal W}[{\cal C}_2]-T_0{\cal W}[{\cal
C}_2^{(0)}]. \label{diff}
\end{equation}
Then, we identify this difference with that computed in the static
gauge since the latter is free from the logarithmic divergence and
is finite in the limit $\varepsilon\to 0$. Since $F(r,0)$ has been
obtained before \cite{chu:2009}, all we need to calculate here is
the difference $F_1(r,T)-F_1(r,0)$.

Parallel to the zero temperature case, we transform the metric
(\ref{ss2}) into conformal one
\begin{equation}
ds^2[{\cal C}_2]=e^{2\phi}(dt^2+d\sigma^2)
\label{2quark}
\end{equation}
in order to remove the artificial singularity at $z=z_0$ introduced by the double coverage of the world sheet ${\cal C}_2$ on the coordinate patch $(t,z)$, where $\sigma$ is related to $z$ via
\begin{equation}
\sigma=\pm\int_z^{z_0}dz\sqrt{\frac{1-f_0}{f(f-f_0)}}. \label{transf}
\end{equation}
We have $\sigma$ goes from -K to 0 and then from 0 to K as z moves from 0 to $z_0$ and back to 0, where
\begin{equation}
K=|\sigma_{max}|=z_0s^2\int_0^1\frac{d\zeta}
{\sqrt{(1-\zeta^4)(s^4-\zeta^4)}}. \label{K}
\end{equation}

Either $z_0$ or $K$ serves a length scale of the problem. The first
two terms of the expansion of z in the powers of $(K\pm\sigma)$
reads
\begin{equation}
z=(K\pm\sigma)[1-\frac{1}{10}(\frac{1}{z_0^4}+\frac{1}{z_h^4})(K\pm\sigma)^4]
\end{equation}
for $K\pm\sigma\ll z_0$.

The conformal factor of (\ref{2quark}) is given by
\begin{equation}
e^{2\phi}=\frac{f}{z^2}
\end{equation}
expressed in terms of $\sigma$ by inverting (\ref{transf}).

In a conformal metric,the scalar curvature R reads
\begin{equation}
R=-2e^{-2\phi}\frac{d^2\phi}{d\sigma^2}=-\frac{2z^2}{f}\frac{d\omega_t^{01}}{d\sigma}
\end{equation}
where $\omega_t^{01}$ is the spin connection
\begin{equation}
\omega_t^{01}=\frac{d\phi}{d\sigma}=-\frac{1}{z}(\frac{2}{f}-1)\frac{dz}{d\sigma}
\end{equation}
To proceed, on writing
\begin{equation}
-\nabla^2\equiv e^{-2\phi}\Delta_0[{\cal C}_2]
\end{equation}
\begin{equation}
-\nabla^2+2+\delta \equiv e^{-2\phi}\Delta_1[{\cal C}_2]
\end{equation}
\begin{equation}
-\nabla^2+4+R-2\delta\equiv e^{-2\phi}\Delta_2[{\cal C}_2]
\end{equation}
\begin{equation}
-i\gamma^\alpha\nabla_\alpha+\tau_3\equiv e^{-\phi}D_F[{\cal C}_2]
\end{equation}
we have
\begin{equation}
\Delta_0[{\cal C}_2]=-\frac{\partial^2}{\partial t^2}-\frac{\partial^2}{\partial\sigma^2},
\label{Delta02}
\end{equation}
\begin{equation}
\Delta_1[{\cal C}_2]=-\frac{\partial^2}{\partial t^2}-\frac{\partial^2}{\partial\sigma^2}
+e^{2\phi}(2+\delta),
\label{Delta12}
\end{equation}
\begin{equation}
\Delta_2[{\cal C}_2]=-\frac{\partial^2}{\partial t^2}-\frac{\partial^2}{\partial\sigma^2}
+e^{2\phi}(4+R-2\delta),
\label{Delta22}
\end{equation}
and
\begin{equation}
D_F[{\cal
C}_2]=-i\left(\frac{\partial}{\partial\sigma}+\frac{1}{2}\frac{d\phi}{d\sigma}
\right)\tau_1-i\frac{\partial}{\partial\tau}\tau_2+e^{\phi}\tau_3
=e^{-\frac{\phi}{2}}\left(-i\tau_1\frac{\partial}{\partial\sigma}
-i\tau_2\frac{\partial}{\partial
t}+\tau_3e^{\phi}\right)e^{\frac{\phi}{2}}. \label{DF2}
\end{equation}
As was shown in \cite{chu:2009}, the scaling factors $e^{-2\phi}$, $e^{-\phi}$ and
$e^{\pm\frac{\phi}{2}}$ do not contribute to the determinant ratio (\ref{parallel}) and
it becomes
\begin{equation}
W[{\cal C}_2]=-\ln\Big[\frac{{\rm
det}^4D_F[{\cal C}_2]}
{{\det}^{\frac{5}{2}}\Delta_0[{\cal C}_2]\,{\det}\Delta_2[{\cal C}_2]\,
{\det}^{\frac{1}{2}}\Delta_1[{\cal C}_2]\,}\Big]
=-\ln\Big[\frac{{\rm det}^2\Delta_+[{\cal C}_2]
{\rm det}^2\Delta_-[{\cal C}_2]}
{{\det}^{\frac{5}{2}}\Delta_0[{\cal C}_2]\,{\det}\Delta_2[{\cal C}_2]\,
{\det}^{\frac{1}{2}}\Delta_1[{\cal C}_2]\,}\Big], \label{parallel1}
\end{equation}
where we have used the relation
\begin{equation}
{\rm det}^2D_F[{\cal C}_2]={\rm det}D_F^2[{\cal C}_2]
={\rm det}\Delta_+[{\cal C}_2]{\rm det}\Delta_-[{\cal C}_2]
\end{equation}
with
\begin{equation}
\Delta_\pm[{\cal C}_2]=-\frac{\partial^2}{\partial t^2}-\frac{\partial^2}{\partial\sigma^2}
+e^{2\phi}\mp e^\phi\frac{d\phi}{d\sigma}.
\end{equation}

Making a Fourier transformation of the time variable $t$, each
functional determinant of (\ref{straight}) and (\ref{parallel}) is
factorized as an infinite product of its Fourier components with
each Fourier component obtained by replacing the time derivative
$\frac{\partial}{\partial\tau}$ in eqs.
(\ref{Delta02})-(\ref{DF2}) by $-i\omega$ with the Matsubara
frequency $\omega=2n\pi T$ for the bosonic determinants and
$\omega=(2n+1)\pi T$ for the fermionic ones. We have then
\begin{equation}
W[{\cal C}_2]=-\sum_{n=-\infty}^\infty\ln\Big[Z_b[2n\pi T]Z_f((2n+1)\pi T)\Big],
\label{summation}
\end{equation}
where
\begin{equation}
Z_b(\omega)={\rm det}D_0^{-\frac{5}{2}}(\omega){\rm det}D_1^{-1}(\omega){\rm det}D_2^{-\frac{1}{2}}(\omega),
\label{boson}
\end{equation}
and
\begin{equation}
Z_f(\omega)={\rm det}D_+^2(\omega){\rm det}D_-^2(\omega)
\label{fermion}
\end{equation}
with
\begin{equation}
D_0(\omega)=-\frac{d^2}{d\sigma^2}+\omega^2,
\label{paralleld0}
\end{equation}
\begin{equation}
D_1(\omega)=-\frac{d^2}{d\sigma^2}+\omega^2+e^{2\phi}(2+\delta),
\label{paralleld1}
\end{equation}
\begin{equation}
D_2(\omega)=-\frac{d^2}{d\sigma^2}+\omega^2+e^{2\phi}(4+R-2\delta),
\label{paralleld2}
\end{equation}
and
\begin{equation}
D_\pm(\omega)=-\frac{d^2}{d\sigma^2}+\omega^2+e^{2\phi}\mp
e^\phi\frac{d\phi}{d\sigma}. \label{paralleldf}
\end{equation}
Notice that $D_+(\omega)\to D_-(\omega)$ upon the reflection $\sigma\to-\sigma$.

The Poisson formula
\begin{equation}
\sum_{n=-\infty}^\infty F(n)=\sum_{m=-\infty}^\infty\int_{-\infty}^\infty dx e^{2im\pi x}F(x)
\end{equation}
enables us to convert the infinite series
(\ref{summation}) into \cite{Hou:2009zk}
\begin{equation}
TW[{\cal C}_2]=-\int_{-\infty}^{\infty}\frac{d\omega}{2\pi}[\ln Z_b(\omega)+\ln Z_f(\omega)]
+\frac{2}{\pi}\int_0^\infty d\varpi\Big[\frac{{\rm Im}\ln Z_b(-i\varpi)}{e^{\beta\varpi}-1}
-\frac{{\rm Im}\ln Z_f(-i\varpi)}{e^{\beta\varpi}+1}\Big].
\end{equation}
$\ln Z_b(\omega)$ and $\ln Z_f(\omega)$ is proportional to the infinite products of
$(\varpi-E)$ with $E$'s the discrete eigenvalues defined by
\begin{equation}
D_\alpha(-iE)\psi=0
\label{eigen}
\end{equation}
for $\alpha=0,1,2,\pm$, subject to the Dirichlet boundary condition
$\psi(\pm K)=0$, and these
eigenvalues are all positive. The integral along the real axis can be carried out readily and we end up with
\begin{eqnarray}
F_1(r,T)-F_1(r,0)&=&TW[{\cal C}_2]-\lim_{T_0\to 0}T_0W[{\cal
C}_2^{(0)}] =-\frac{1}{i\pi}\int_{0}^{i\infty}d\omega
\ln\frac{Z_b(\omega)Z_f(\omega)}{Z_b^{(0)}(\omega)Z_f^{(0)}(\omega)}\\\nonumber
&-&\frac{4}{\beta}\Big[-\frac{5}{2}\sum_{E_0}\ln(1-e^{-\beta
E_0})-\sum_{E_1}\ln(1-e^{-\beta
E_1})-\frac{1}{2}\sum_{E_2}\ln(1-e^{-\beta
E_2})+4\sum_{E_3}\ln(1+e^{-\beta E_3})\Big] \label{f1-f10}
\end{eqnarray}
where $E_k$ ($k=0,1,2$) stands for the common eigenvalues of $D_k(\omega)$ while $E_3$ means the common eigenvalues of $D_\pm(\omega)$ and the superscript "(0)"
denote the corresponding quantities at $T=0$. The dimensionless function
$g_1(rT)$ in (\ref{expansion}) is given by
\begin{equation}
g_1(rT)=-\frac{\Gamma^4{(\frac{1}{4})}r}{4\pi^2\kappa}[F_1(r,T)-F_1(r,0)]
\label{g1}
\end{equation}
with $\kappa=-1.33460...$ and its numerical values of the function $g_1(rT)$ will be computed in the next section.

\section{The evaluation of $g_1(rT)$}

It follows from (\ref{boson}) and (\ref{fermion}) that the quantity
$Z_b(\omega)Z_f(\omega)$ contains equal powers of functional
determinants in the numerator and the denominator and the problem
boils down to the evaluation of determinant ratios, which can be
tackled with the Gelfand-Yaglom method
\cite{Gelfand}\cite{Kleinert}.

Consider two functional operators
\begin{equation}
H_\alpha = -\frac{d^2}{dx^2}+V_\alpha(x),
\end{equation}
with $\alpha=1,2$, defined in the domain $a\le x\le b$ under the Dirichlet boundary condition, it was
shown that the determinant ratio
\begin{equation}
\frac{{\rm det}H_2}{{\rm det}H_1}=\frac{f_2(b|a)}{f_1(b|a)}
\label{dratio}
\end{equation}
where $f_\alpha(x|a)$ is the solution of the homogeneous equation
\begin{equation}
H_\alpha f=0, \label{homo}
\end{equation}
subject to the conditions $f_\alpha(0|a)=0$ and
$f_\alpha^\prime(0|a)=1$. In terms of a pair of linearly independent
solutions of (\ref{homo}), $(\eta_\alpha,\xi_\alpha)$
\begin{equation}
f_\alpha(b|a)=\frac{\eta_\alpha(a)\xi_\alpha(b)-\eta_\alpha(b)\xi_\alpha(a)}
{W[\eta_\alpha,\xi_\alpha]}.
\label{lambda}
\end{equation}
where the Wronskian $W[\eta_\alpha,\xi_\alpha]$ is $x$-independent.
This method has been employed recently in
\cite{Kruczenski:2008zk} to calculate the one loop effective action
of a Wilson line or that of a circular Wilson loop. See \cite{Dunne}
for a review on other applications.

Coming back to the subleading term of the heavy quark
potential, the operator $H_\alpha$ corresponds to one of the
operators (\ref{paralleld0})$\sim$(\ref{paralleldf}). With $H_\alpha$ given by the operator
(\ref{paralleld0}), eq.(\ref{homo}) can be solved analytically and
we may choose the following pairs of independent solutions
\begin{equation}
\eta_0=\sinh\omega\sigma \qquad \xi_0=\cosh\omega\sigma.
\label{etaxi0}
\end{equation}
with their Wronskian is given by
\begin{equation}
W[\eta_0,\xi_0]=-\omega,
\end{equation}

The equations (\ref{homo}) with $H_\alpha$ given by
(\ref{paralleld1}), (\ref{paralleld2}) and (\ref{paralleldf})
\begin{equation}
D_1(\omega)\psi=
\frac{d^2\psi}{d\sigma^2}-[\omega^2+e^{2\phi}(2+\delta)]\psi=0,
\label{potential1}
\end{equation}
\begin{equation}
D_2(\omega)\psi=
\frac{d^2\psi}{d\sigma^2}-[\omega^2+e^{2\phi}(4+R-2\delta)]\psi=0,
\label{potential2}
\end{equation}
and
\begin{equation}
D_\pm(\omega)\psi= \frac{d^2\psi}{d\sigma^2}-[\omega^2+e^{2\phi}\mp
e^\phi\frac{d\phi}{d\sigma}]\psi=0. \label{potential+}
\end{equation}
do not admit analytical solutions and we shall express the quantity of $f$
\label{lambda} for each case of (\ref{Delta02})-(\ref{DF2}) in terms of the
Wronskian between a pair of special solutions. Because of the asymptotic
behavior $e^{2\phi}\simeq\frac{1}{z^2}$, $\delta\simeq 0$ and $R\simeq-2$
for $\sigma$ close to $\pm K$,
eqs.(\ref{potential1}) and (\ref{potential2}) have $\sigma=\pm K$ as
regular points with the same pair of indexes (2,-1). The equation
$D_\pm(\omega)\phi=0$ has a regular point $\sigma=\mp K$ with the
indexes (2,-1) and $\sigma=\pm K$ is an ordinary point of it. We associate
$\eta_\alpha$ to the vanishing solution at $\sigma=-K$ and
$\xi_\alpha$ to the vanishing solution at $\sigma=K$ with the
normalization conditions
\begin{equation}
\lim_{\sigma\to -K}\frac{\eta_1(\sigma)}{\omega^2(\sigma+K)^2}
=\lim_{\sigma\to -K}\frac{\eta_2(\sigma)}{\omega^2(\sigma+K)^2}
=\lim_{\sigma\to -K}\frac{\eta_+(\sigma)}{\omega^2(\sigma+K)^2}=1
\label{etadef}
\end{equation}
and
\begin{equation}
\lim_{\sigma\to K}\frac{\xi_1(\sigma)}{\omega^2(K-\sigma)^2}
=\lim_{\sigma\to K}\frac{\xi_2(\sigma)}{\omega^2(K-\sigma)^2}
=\lim_{\sigma\to K}\frac{\xi_-(\sigma)}{\omega^2(K-\sigma)^2}=1.
\label{xidef}
\end{equation}
Furthermore, we require
\begin{equation}
\eta_-(-K)=\xi_+(K)=0
\label{cond1}
\end{equation}
and
\begin{equation}
\eta_-^\prime(-K)=-\xi_+^\prime(K)=\omega
\label{cond2}
\end{equation}
with the prime the derivative with respect to $\sigma$.
On account of the eveness of $D_1(\omega)$ and $D_2(\omega)$ with respect to $\sigma$,
we have
\begin{equation}
\xi_{1,2}(\sigma)=\eta_{1,2}(-\sigma).
\label{symmetry1}
\end{equation}
It follows from the relation between $D_+(\omega)$ and $D_-(\omega)$ that
\begin{equation}
\eta_-(\sigma)=\xi_+(-\sigma) \qquad \xi_-(\sigma)=\eta_+(-\sigma)
\label{symmetry2}
\end{equation}
Each differential equation of (\ref{potential1}), (\ref{potential2})
and (\ref{potential+}) is of the form of a one dimensional
Schrodinger equation at zero energy and does not admit a bound state
subject to the Dirichlet boundary condition. Therefore we expect
that
\begin{equation}
\eta_{1,2}(\sigma)=\frac{C_{1,2}(\omega)}{\omega(K-\sigma)}+...,
\end{equation}
\begin{equation}
\eta_-(\sigma)=\frac{C_-(\omega)}{\omega(K-\sigma)}+...,
\end{equation}
as $\sigma\to K$ and
\begin{equation}
\xi_{1,2}(\sigma)=\frac{C_{1,2}(\omega)}{\omega(K+\sigma)}+...,
\end{equation}
\begin{equation}
\xi_+(\sigma)=\frac{C_+(\omega)}{\omega(K+\sigma)}+...,
\end{equation}
as $\sigma\to -K$. The coefficients of divergence, $C_1(\omega)$,
$C_2(\omega)$ and $C_\pm(\omega)$ are nonzero for $\omega\ne 0$ and are related to the Wronskian's via
\begin{equation}
C_j(\omega)=-\frac{W[\eta_j,\xi_j]}{3\omega}
\label{CtoW}
\end{equation}
with $j=1,2,\pm$. We have, in addition
\begin{equation}
\eta_+(K)=-\frac{W[\eta_+,\xi_+]}{\omega}
\qquad
\xi_-(-K)=-\frac{W[\eta_-,\xi_-]}{\omega}
\label{alternative}
\end{equation}
It follows from the symmetry property (\ref{symmetry2}) that
\begin{equation}
C_+(\omega)=C_-(\omega)\equiv C_3(\omega)
\end{equation}
For $\omega>>1/K$, the solutions $\eta$'s and $\xi$'s can be
approximated by WKB method and we find the asymptotic behaviors
\begin{equation}
C_1(\omega)=\frac{3}{2}e^{{2K\omega}+\frac{I+J}{\omega}+...},
\label{largeomega1}
\end{equation}
\begin{equation}
C_2(\omega)=\frac{3}{2}e^{{2K\omega}+\frac{2(I-J)}{\omega}+...},
\label{largeomega2}
\end{equation}
and
\begin{equation}
C_3(\omega)=\frac{1}{2}e^{{2K\omega}+\frac{I}{2\omega}+...},
\label{largeomega3}
\end{equation}
where
\begin{equation}
I\equiv\int_0^{z_0}dz\sqrt{g}(R+2)=\frac{2}{z_0s^2}\int_0^1d\zeta\zeta^2\frac{-s^8+4s^4-6s^4\zeta^4+3\zeta^8}
{\sqrt{(1-\zeta^4)(s^4-\zeta^4)^3}}
\end{equation}
and
\begin{equation}
J\equiv\int_0^{z_0}dz\sqrt{g}\delta=\frac{2}{z_0s^2}\int_0^1d\zeta\zeta^2\sqrt{\frac{1-\zeta^4}{s^4-\zeta^4}}.
\end{equation}
The details of the derivation are shown in the appendix A. These
asymptotic forms guarantees the UV convergence of the integral in
(\ref{F_1(r,T)-F_1(r,0)}) below. Also they will be used to check the
numerical solutions. Here we present four curves about
(\ref{largeomega1}),(\ref{largeomega2}) and (\ref{largeomega3}) at
two different temperatures represented by the quantity s  (For $s=\frac{z_h}{z_0}=\frac{1}{T\pi z_0}$) in Figure 1.
\begin{figure}
\centering
\includegraphics[width=0.41\textwidth]{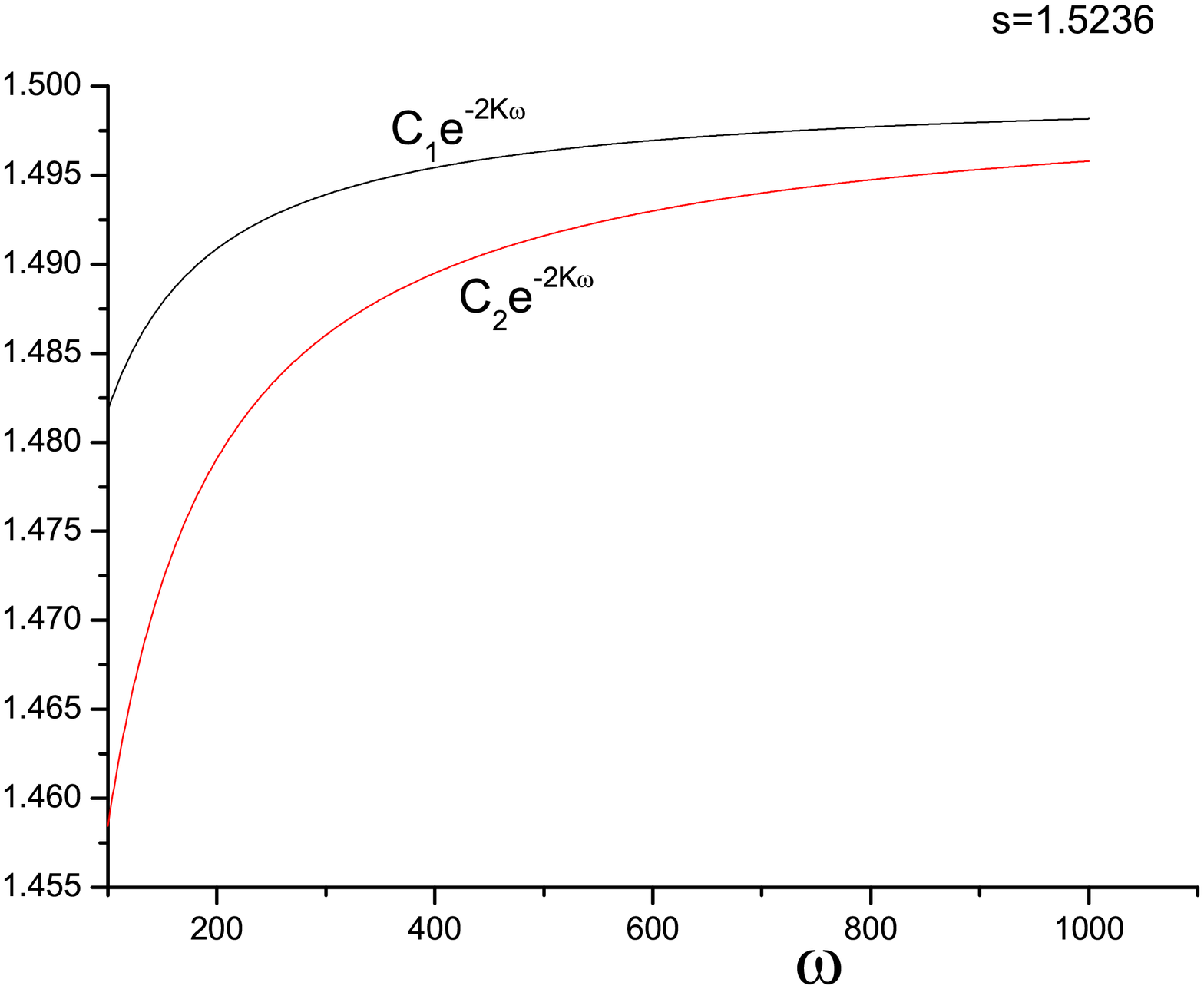}
\includegraphics[width=0.41\textwidth]{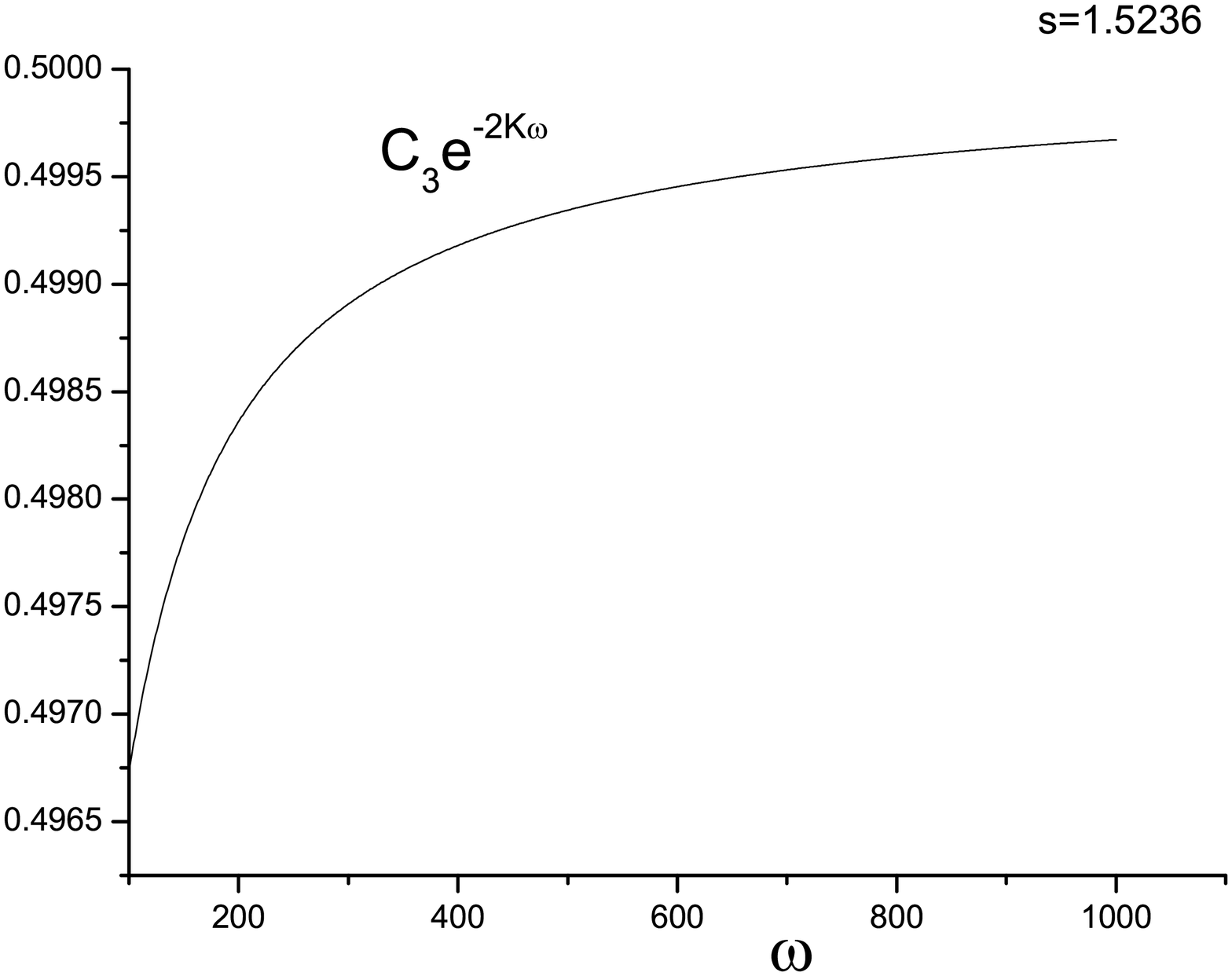}
\includegraphics[width=0.41\textwidth]{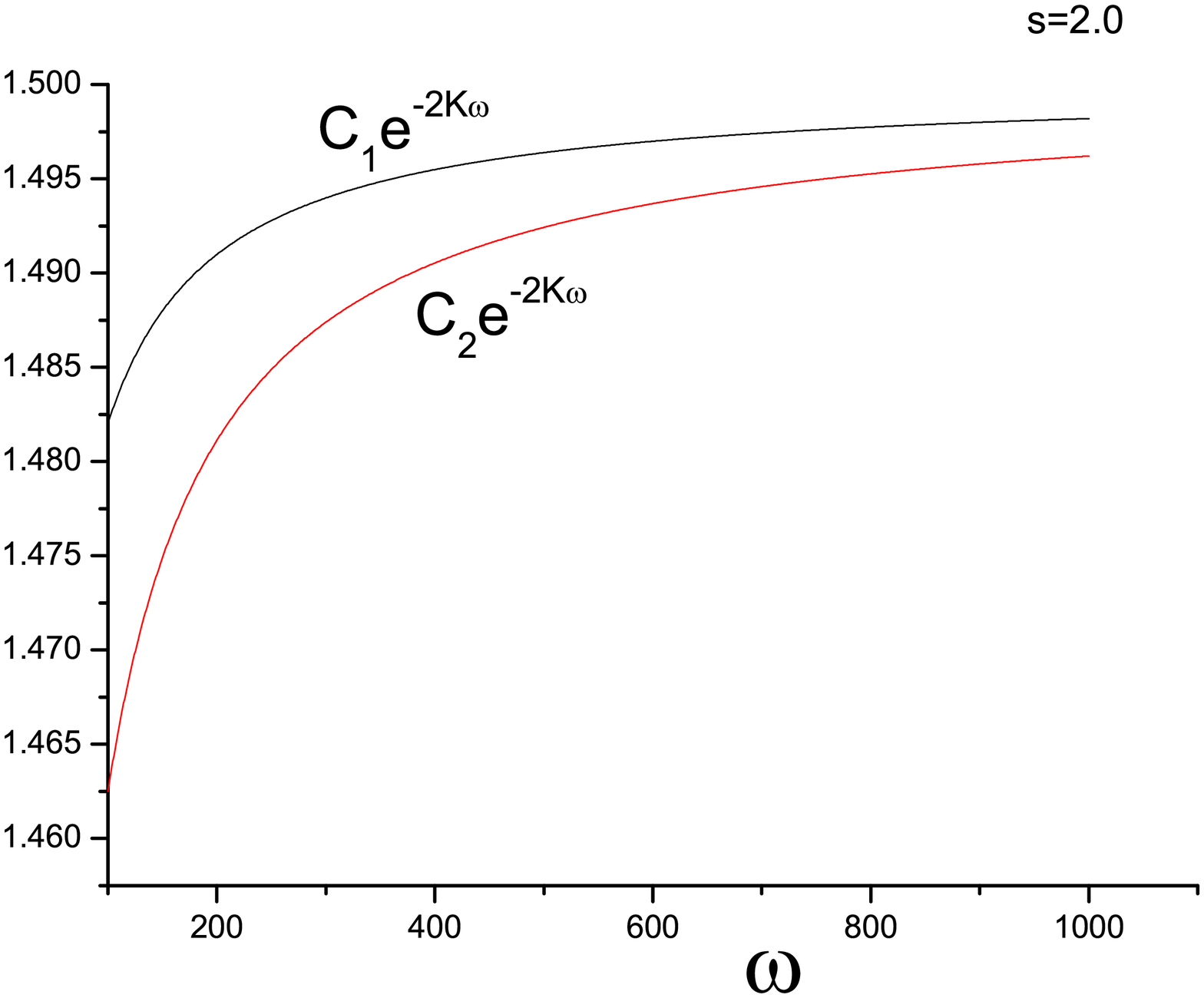}
\includegraphics[width=0.41\textwidth]{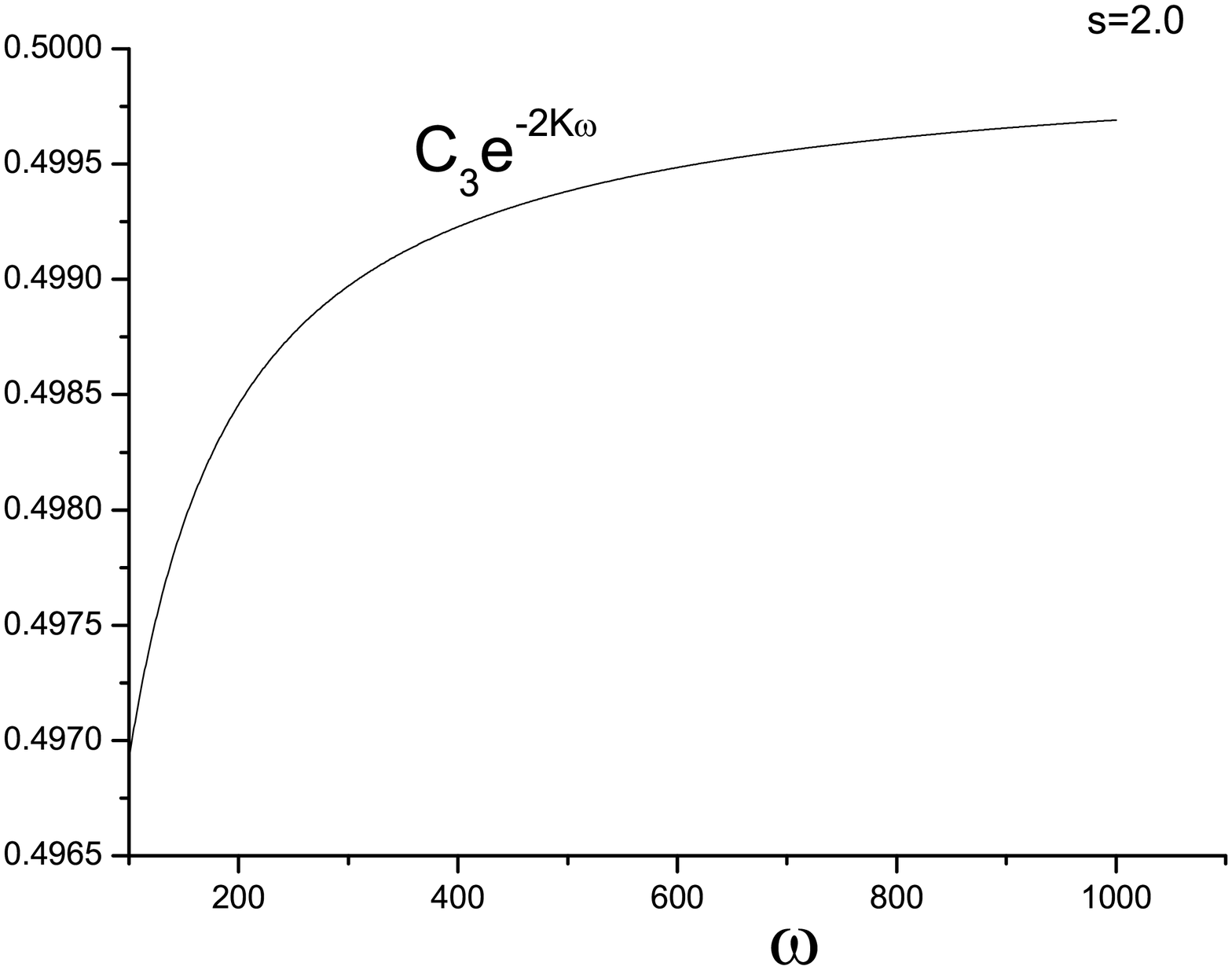}
\caption{The top two curves represent $C_1e^{-2K\omega}$,
$C_2e^{-2K\omega}$ and $C_3e^{-2K\omega}$ at s=1.5236, the bottom
represent the same quantities that at s=2.0.}
\end{figure}

The small $\omega$ behavior can be obtained by introducing an
another set of solutions, normalized differently

\begin{equation}
\bar\eta_{1,2,+}(\sigma)\equiv\frac{\eta_{1,2,+}(\sigma)}{\omega^2}, \qquad
\bar\xi_{1,2,-}(\sigma)\equiv\frac{\xi_{1,2,-}(\sigma)}{\omega^2}
\end{equation}
and
\begin{equation}
\bar\eta_-(\sigma)\equiv\frac{\eta_-(\sigma)}{\omega}
\qquad \bar\xi_+(\sigma)\equiv\frac{\xi_+(\sigma)}{\omega}.
\end{equation}
Defining the coefficients
$\bar C$'s by the diverging behavior
\begin{equation}
\bar\eta_{1,2,-}(\sigma)=\frac{\bar C_{1,2,-}(\omega)}{K-\sigma}+...,
\end{equation}
as $\sigma\to K$ and
\begin{equation}
\bar\xi_{1,2,+}(\sigma)=\frac{\bar C_{1,2,+}(\omega)}{K+\sigma}+...,
\end{equation}
as $\sigma\to-K$, we find that $C_{1,2}(\omega)\sim\omega^3$ and
$C_3(\omega)\sim\omega^2$ as $\omega\to 0$.

Now we apply the Gelfand-Yaglom method with
$-K+\epsilon\le\sigma\le\,K-\epsilon$ and designate
$U_\alpha(\omega)$ to the quantity $f$ of (\ref{lambda}) with
$\alpha=0,1,2,\pm$ corresponding to the indexes of the operators
(\ref{paralleld0})-(\ref{paralleldf}), and the superscript "(0)"
denote the corresponding quantities at $T=0$. We find that
%
\begin{equation}
{\cal R}_2(\omega)=\frac{U_+^2(\omega)U_-^2(\omega)}
{U_0^{\frac{5}{2}}(\omega)U_1(\omega)U_2^{\frac{1}{2}}(\omega)}
\label{R2}
\end{equation}
and
\begin{equation}
{\cal
R}_2^{(0)}(\omega)=\frac{{[U_+^{(0)}(\omega)]}^2{[U_-^{(0)}(\omega)]}^2}
{[U_0^{(0)}(\omega)]^{\frac{5}{2}}U_1^{(0)}(\omega)[U_2^{(0)}(\omega)]^{\frac{1}{2}}}.
\label{R2^{(0)}}
\end{equation}
where the superscript "(0)" denote the corresponding quantities at $T=0$.
It follows from (\ref{etaxi0}) that
\begin{equation}
U_0(\omega)=\frac{\sinh2(K-\epsilon)\omega}{\omega}.
\label{U0}
\end{equation}
The symmetry (\ref{symmetry2}) implies that
$U_+(\omega)=U_-(\omega)$. In the limit $\epsilon\to 0$

\begin{equation}
U_{1,2}(\omega)\simeq\frac{C_{1,2}(\omega)}{3\omega^3\epsilon^2}
\label{approx2}
\end{equation}
and
\begin{equation}
U_\pm(\omega)\simeq\frac{C_3(\omega)}{\omega^2\epsilon}.
\label{approx3}
\end{equation}
The counterparts at $T=0$ have the similar formula. Substituting
(\ref{U0})$\sim$(\ref{approx3}) into (\ref{f1-f10}),we have
\begin{eqnarray}
&{}&F_1(r,T)-F_1(r,0)=-\frac{1}{\pi}\int_{0}^{\infty}d\omega
\ln\frac{C^4_3C^{(0)}_1{C^{(0)}_2}^\frac{1}{2}{\sinh^{(0)}}^\frac{5}{2}(2K^{(0)}\omega)}{{C^{(0)}_3}^4C_1{C_2}^\frac{1}{2}{\sinh}^\frac{5}{2}(2K\omega)}\nonumber\\
&-&\frac{4}{\beta}(-\frac{5}{2}\sum_{E_0}\ln(1-e^{-\beta
E_0})-\sum_{E_1}\ln(1-e^{-\beta
E_1})-\frac{1}{2}\sum_{E_2}\ln(1-e^{-\beta
E_2})+4\sum_{E_3}\ln(1+e^{-\beta E_3})), \label{F_1(r,T)-F_1(r,0)}
\end{eqnarray}
The first term of (\ref{F_1(r,T)-F_1(r,0)}) depends on
$C_1,C_2,C_3$ and $C_1^{(0)},C_2^{(0)},C_3^{(0)}$, which can be
obtained by solving the Schrodinger like equations
(\ref{potential1})$\sim$(\ref{potential+}). It follows from
(\ref{largeomega1}), (\ref{largeomega2}) and (\ref{largeomega3})
that the integral converges. Notice that the temperature dependence
is entirely through $s$, so we can use the same method to deal with
$C_1,C_2,C_3$ and $C_1^{(0)},C_2^{(0)},C_3^{(0)}$. The potentials of
the Schrodinger like equations,
$e^{2\phi}(2+\delta), e^{2\phi}(4+R-2\delta)$ and $e^{2\phi}\mp
e^\phi\frac{d\phi}{d\sigma}$ are elementary functions of $z$ but
depend $\sigma$ via (\ref{transf}). So at the beginning, we use the
forth order Runge-Kutta method(RK4) to the first order differential
equation
\begin{equation}
\frac{dz}{d\sigma}=\pm\sqrt{\frac{f(f-f_0)}{1-f_0}}
\end{equation}
to find the inverse of (\ref{transf}), which maps the equal step
length with respect to $\sigma\in[-K+\epsilon,K]$ and to unequal step
length with respect to $z\in[\epsilon,z_0]$ according to
(\ref{transf}), then again use RK4 with the boundary conditions
(\ref{etadef}) and (\ref{cond2}) to solve these equations. Here we
present two curves about z versus $\sigma$ at two different
temperatures in Figure 2.
\begin{figure}[h]
\centering
\includegraphics[width=0.41\textwidth]{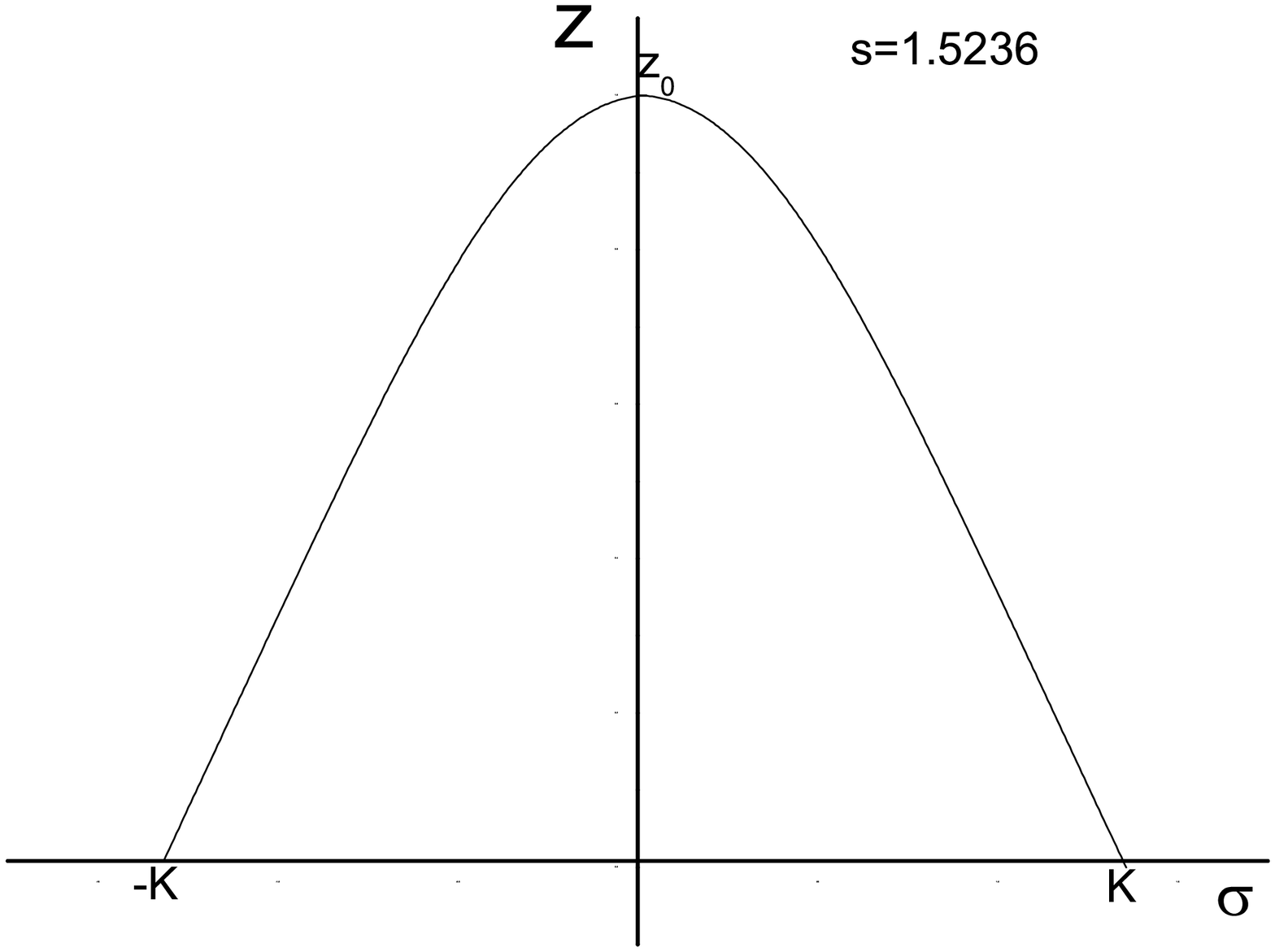}
\includegraphics[width=0.41\textwidth]{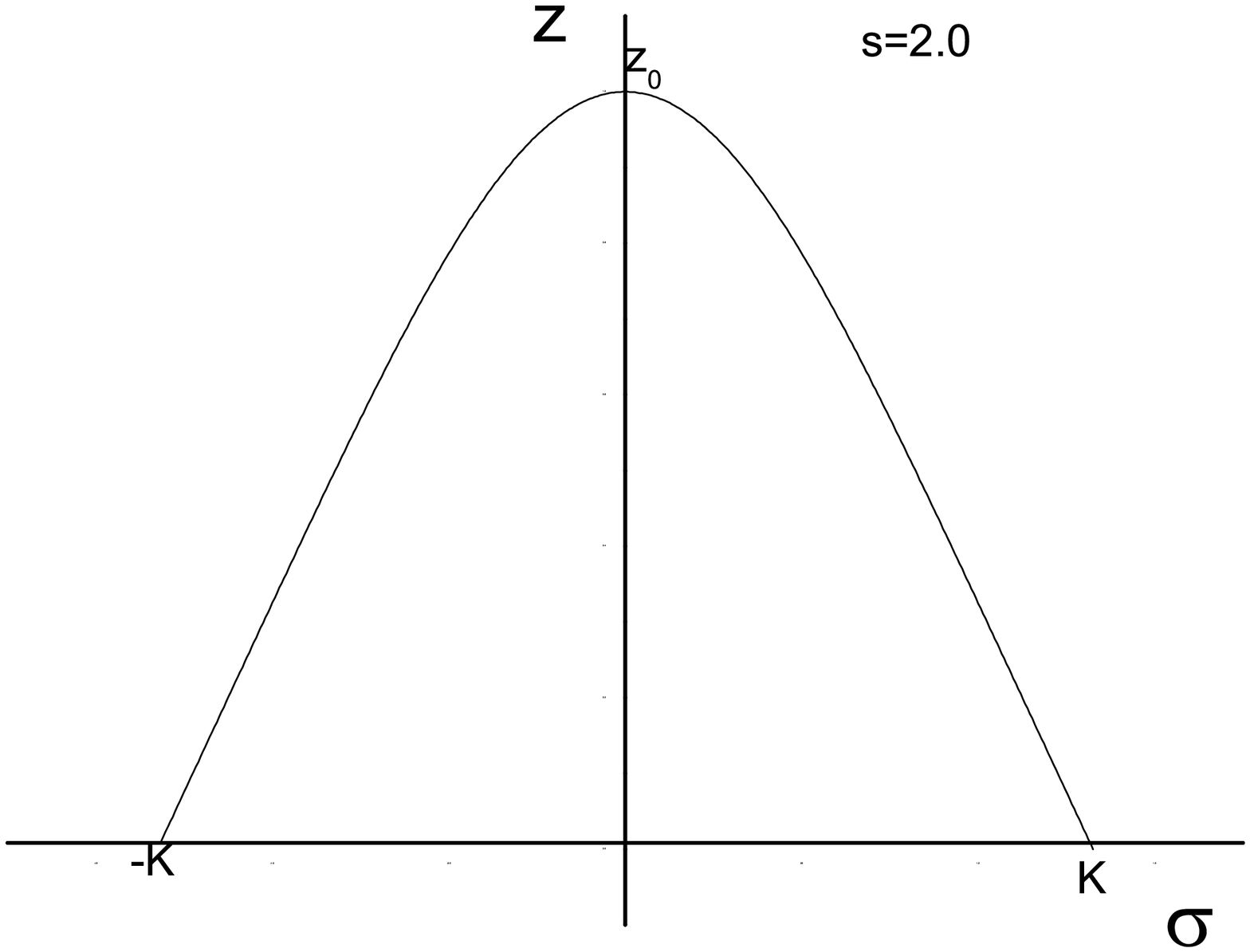}
\caption{The left corresponds to z versus $\sigma$ at s=1.5236 with
$K\cong1.356203z_0$,  the right corresponds to z versus $\sigma$ at
s=2.0 with $K\cong1.325163z_0$.}
\end{figure}

For eqs.(\ref{potential1}) and (\ref{potential2}), we take advantage
of the symmetry property (\ref{symmetry1}) and evaluate the
Wronskian by the formula
\begin{equation}
W_{1,2}(\omega)=-2\eta_{1,2}(0)\eta_{1,2}^\prime(0)
\label{formulaW}
\end{equation}
The coefficients $C_{1,2}(\omega)$ follow from (\ref{CtoW}). For
eq.(\ref{potential+}) with the upper sign, we develop both
$\eta_+(\sigma)$ from $\sigma\simeq -K$ and $\xi_+(\sigma)$ from
$\sigma=K$, evaluate their Wronskian at $\sigma=0$ and calculate the
coefficient $C_3(\omega)$ from eq.(\ref{CtoW}). An alternative way
is to run the solution $\eta_+(\sigma)$ all the way to $K$ and
calculate the Wronskian by eq.(\ref{alternative}). To avoid the
rapid changes of the potential function near the singularity
$\sigma=-K$, we start with an analytical approximation of
$\eta_{1,2}(\sigma)$ and $\eta_+(\sigma)$ at $\sigma=-K+\rho$ with
$\epsilon<<\rho<<K$ and then run the Runge-Kutta iteration for
$\sigma>-K+\rho$. On writing $x=\omega\rho$, we find the approximate
solutions
\begin{equation}
\eta_{1,2}(-K+\delta)=3\lbrace u_1(x)+c_{1,2}[p(x)u_1(x)+q(x)v_1(x)]\rbrace
\end{equation}
and
\begin{equation}
\eta_+(-K+\delta)=3\lbrace u_1(x)+c_+[p(x)u_1(x)+q(x)v_1(x)]\rbrace
\end{equation}
where
\begin{equation}
u_1(\omega\sigma)=\cosh\omega\sigma-\frac{\sinh\omega\sigma}{\omega\sigma}\qquad
v_1(\omega\sigma)=\left(1+\frac{1}{\omega\sigma}\right)e^{-\omega\sigma}, \label{uv1}
\end{equation}
\begin{equation}
p(x)=\frac{1}{20\omega^4}\Big[\frac{1}{3}x^3-x+\frac{5}{4}
-\left(\frac{1}{2}x^2+\frac{3}{2}x+\frac{5}{4}\right)e^{-2x}\Big]
\end{equation}
and
\begin{equation}
q(x)=-\frac{1}{20\omega^4}\Big[\frac{1}{3}x^3-x+\frac{1}{2}\left(x^2
+\frac{5}{2}\right)\sinh 2x-\frac{3}{2}x\cosh 2x\Big].
\end{equation}
The coefficients $c_1=4+\frac{4}{s^4}$, $c_2=-16-\frac{56}{s^4}$ and
$c_+=-1-\frac{6}{s^4}$. No such a precaution is necessary for the
solution $\xi_+(\sigma)$ and the Runge-Kutta can start right at the
point $\sigma=K$.
The second term of (\ref{F_1(r,T)-F_1(r,0)}) is dependent on
${E_\alpha}$, which refers to the eigenvalues of $D_\alpha(-iE)$
defined by (\ref{eigen}) and they are all positive. For $D_0(-iE)$,
we find
\begin{equation}
E_0=\frac{n\pi}{2K}
\end{equation}
with $n$ positive integers. The eigenvalues pertaining to other $D_\alpha$'s
satisfies the condition
\begin{equation}
C_\alpha(-iE)=0.
\end{equation}
Replacing $\omega^2$ to $-E^2$ in eqs. (\ref{potential1}),
(\ref{potential2}) and (\ref{potential+}), we can obtain the
coefficients $C_\alpha(-iE)$ with the same RK4 method and thereby
its zeros. Some low-lying eigenvalues of each $D_\alpha(-iE)$ at
different temperatures  are presented in the Table 1.
\begin{table}[h]
\centering
\begin{tabular}{|c|c|c|c|c|c|c|}\hline
s&$E_{11}$&$E_{12}$&$E_{21}$&$E_{22}$&$E_{31}$&$E_{32}$ \\ \hline
1.5236&2.10070&3.34648&1.47031&3.13753&1.55800&2.80913\\\hline
1.55&2.10802&3.35507&1.49547&3.14910&1.56451&2.81671\\\hline
1.60&2.12013&3.36926&1.53651&3.16823&1.57527&2.82923\\\hline
1.65&2.13032&3.38118&1.57056&3.18433&1.58435&2.83975\\\hline
1.66&2.13216&3.38333&1.57666&3.18725&1.58599&2.84167\\\hline
1.67&2.13395&3.38541&1.58256&3.19006&1.58758&2.84350\\\hline
1.68&2.13568&3.38743&1.58826&3.19279&1.58912&2.84528\\\hline
1.70&2.13897&3.39127&1.59909&3.19799&1.59206&2.84868\\\hline
1.80&2.15270&3.40726&1.64378&3.21962&1.60430&2.86282\\\hline
2.00&2.17068&3.42816&1.70129&3.24794&1.62036&2.88132\\\hline
2.50&2.19042&3.45103&1.76318&3.27901&1.63802&2.90159\\\hline
10.00&2.20378&3.46646&1.80438&3.29997&1.64998&2.91527\\\hline
200.00&2.20445&3.46652&1.80621&3.30005&1.65214&2.91533\\\hline
\multicolumn{3}{c}{}
\end{tabular}
\label{exampletable} \caption{The first and second eigenvalue of each $D_\alpha(-iE)$ at
different s(a measure of temperature). Where $E_{\alpha1}$ stands for the first eigenvalue of $D_\alpha$ and $E_{\alpha2}$ means the second eigenvalue of $D_\alpha$. }
\end{table}

We find that the spacing between successive eigenvalues converges quickly to the
asymptotic value $\frac{\pi}{2K}$ for large eigenvalues. Furthermore
the eigenvalues hardly
change when $s>2.5$. Such a limiting behavior implies above
the critical s, they almost return to the case of zero temperature.

In the last step,  substituting (\ref{F_1(r,T)-F_1(r,0)}) into
(\ref{g1}), we finally obtain the numerical values of $g_1(rT)$ and
the correction to the screening radius. Figure 3 is the curve of
$g_1(rT)$ versus $rT$. The curve of $g_0(rT)$ is also displayed for
reference. The physical meaning of $g_1(rT)$ will be discussed in
the next section.

\begin{figure}[h]
\centering
\includegraphics[width=0.47\textwidth]{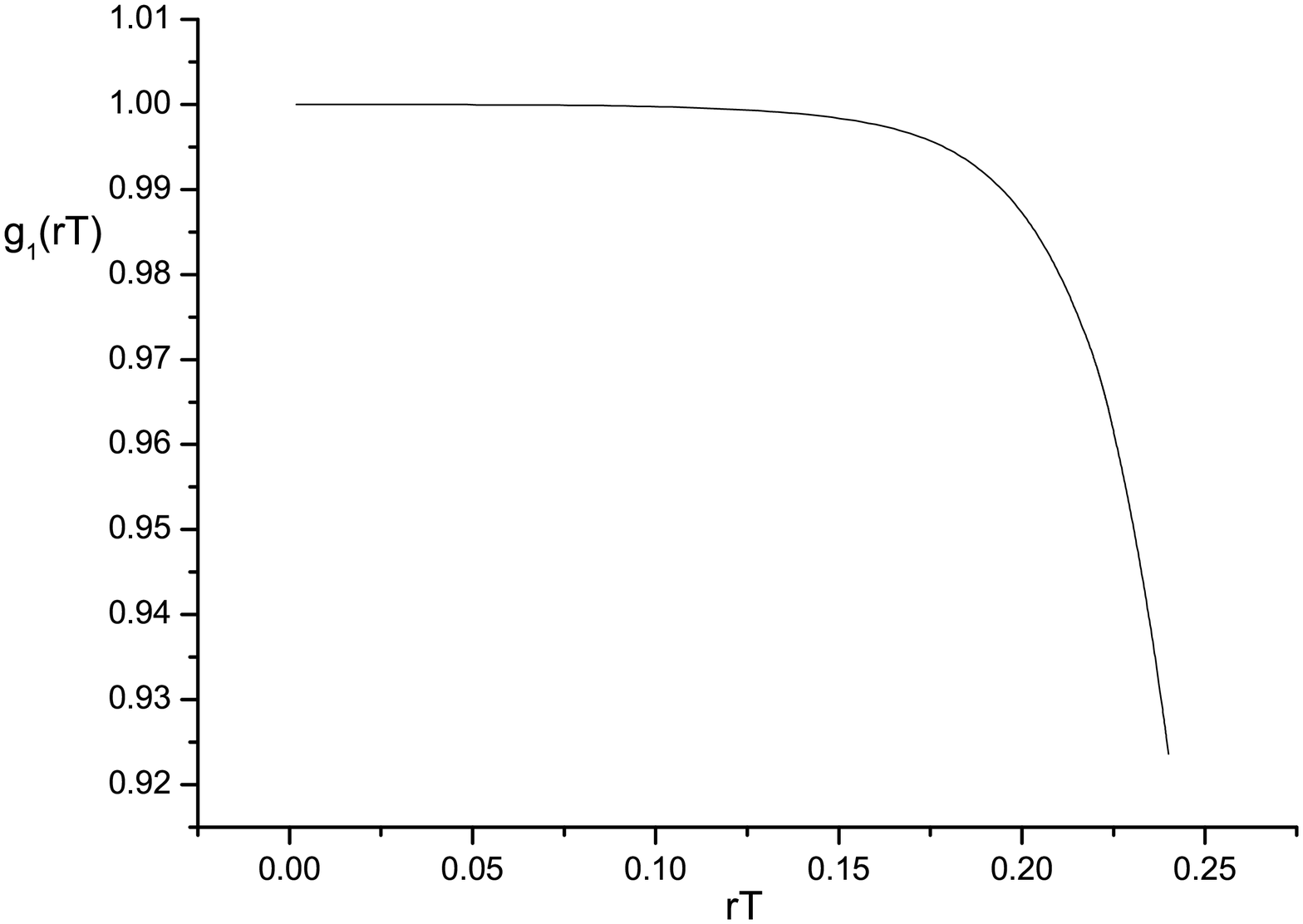}
\includegraphics[width=0.47\textwidth]{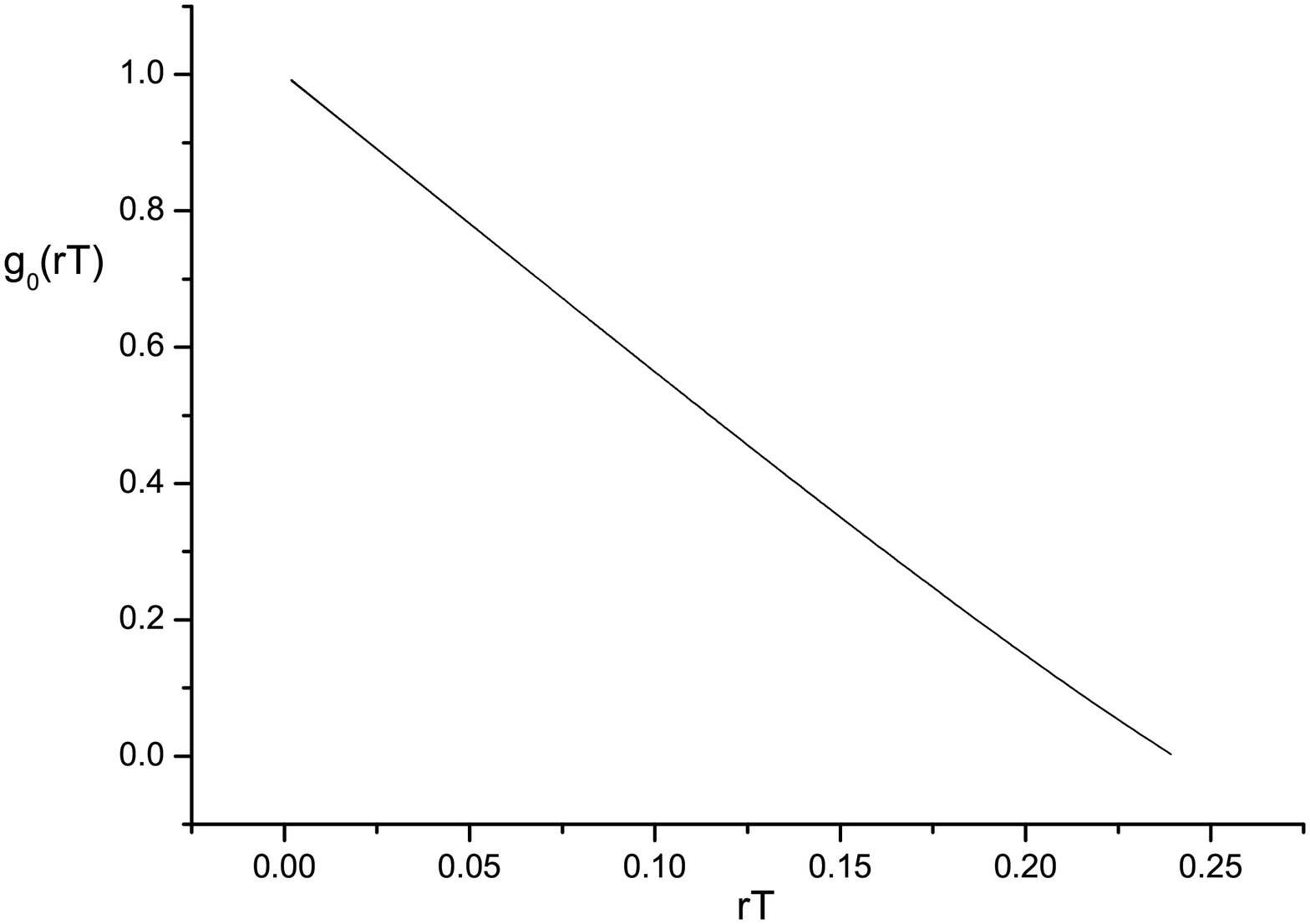}
\caption{The left curve represents $g_1(rT)$, while the right
represents $g_0(rT)$.}
\end{figure}

\section{Concluding remarks}


In this paper, we generalized the previous calculation of the order
$O\left(\frac{1}{\sqrt{\lambda}}\right)$ correction of the heavy
quark potential at $T=0$ to the case at $T>0$. We start from the
partition function of the fluctuation around the world sheet
underlying the leading order heavy quark potential with a Euclidean
time $0<t<1/T$ and calculate its logarithm, the one loop effective
action. The partition function is a product of ratios of the
functional determinants of the world sheet Laplacian plus mass terms.
Each determinant ratio is then decomposed into an infinite product
over Matsubara energies with each factor a ratio of one-dimensional
functional determinants. The summation over Matsubara energies in
the effective action is converted into an integral over Euclidean
energies and an infinite series over the eigenvalues of the
functional operators involved. The determinant ratios in the
integrand are evaluated numerically by the Gelfand-Yaglom method and
the eigenvalues are determined similarly.

Like the zero temperature case, the order
$O\left(\frac{1}{\sqrt{\lambda}}\right)$ correction reduces the
magnitude of the heavy quark potential and leads to a smaller
screening radius. The magnitude of reduction, measured by $\kappa
g_1(rT)$, follows closely that at zero temperature within the
screening radius. The function $g_1(rT)$ decreases from one at
$rT=0$ to about 0.92 at the screening radius. Therefore, within ten
percent of accuracy, the subleading correction may be approximated
by that at $T=0$, i.e.
\begin{equation}
\delta F(r,T)=\frac{4\pi^2}{\Gamma^4\left(\frac{1}{4}\right)}\times\frac{1.3346}{r}
\end{equation}

In the potential model of the heavy quarkonia, the order $O\left(\frac{1}{\sqrt{\lambda}}\right)$ correction will lower the
meson melting temperature.

In the weak coupling, the heavy quark potential at $T>0$ is of
Yukawa type, that is nonvanishing for arbitrarily large $r$. To the
leading order of strong coupling, on the other hand, the magnitude
of the potential drops to zero at a finite $r$. For a smooth cross
over from the strong coupling to the weak coupling, one would expect
the order $O\left(\frac{1}{\sqrt{\lambda}}\right)$ term increases
the screening radius. This would require the function $g_1(rT)$
vanishes somewhere within the screening radius. This, however, is
not the case  as  shown  in Figure 4. Therefore, higher order terms are
required to reverse the trend when $\lambda$ is lowered
sufficiently. Because of the continuity of the free energy $F(r,T)$
with respect to its arguments and the coupling, it is expected that
the cutoff nature of the potential, $F(r,T)=0$ for $r>r_0$, remains
in the weak coupling regime with $r_0$ much longer than the Debye
length of the Yukawa potential.

\begin{figure}[h]
\centering
\includegraphics[width=0.43\textwidth]{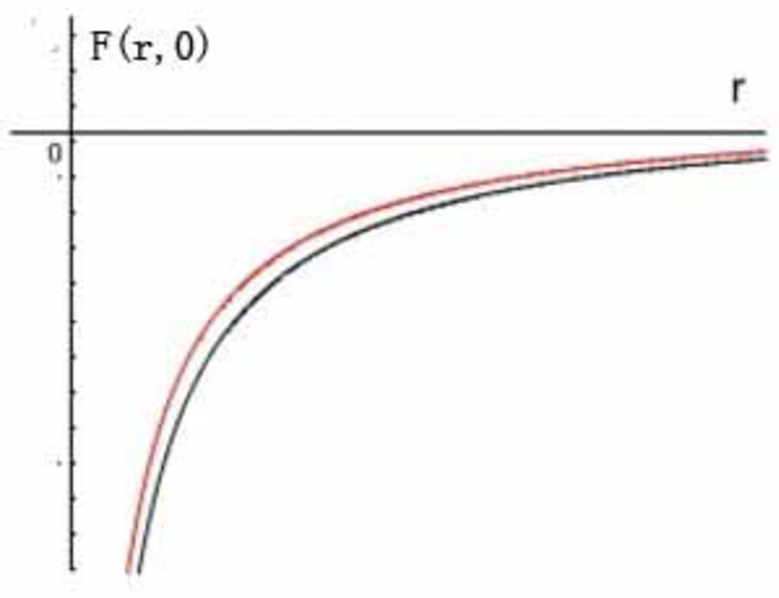}
\includegraphics[width=0.43\textwidth]{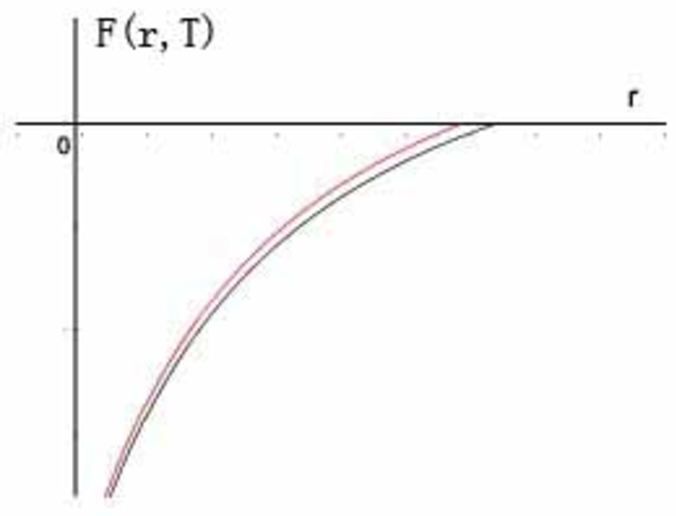}
\caption{The left figure represents  heavy quark potentials at zero
temperature, while the right one represents the potentials at  finite temperature. In both figures, the top lines correspond to the potential including
subleading order correction and the bottom ones  are leading order results.}
\end{figure}

\acknowledgments

The research of D. F. H. and H. C. R. is supported in part
by NSFC under grant Nos. 10575043, 10735040.   D.F. H would also like to take this opportunity to thank KITPC at Beijing for hosting an enjoyable
program entitled AdS/CFT and Novel Approaches to Hadron and Heavy Ion Physics (10/11 - 12/03, 2010), during which part of this manuscript was written.


\appendix

\section {}

In this appendix, we present the details of the derivation of
$C_1,C_2,C_3$ at large $\omega$.\\
\subsection {Bosonic contribution}
We start with the 2nd order ordinary differential equations
\\
\begin{equation}
[-\frac{d^2}{d\sigma^2}+\omega^2+e^{2\phi}(2+\delta)]\eta_1=0
\end{equation}
and
\begin{equation}
[-\frac{d^2}{d\sigma^2}+\omega^2+e^{2\phi}(4+R-2\delta)]\eta_2=0
\end{equation}
As $\sigma\rightarrow-K$, denote $\sigma=-K+x$ with $x\ll1$, we have
\begin{equation}
(-\frac{d^2}{dx^2}+\omega^2+\frac{2}{x^2})\eta_1=0
\end{equation}
\begin{equation}
(-\frac{d^2}{dx^2}+\omega^2+\frac{2}{x^2})\eta_2=0
\end{equation}
The vanishing solution at $\sigma=-K$ reads
\begin{equation}
\eta_1\cong\eta_2\cong3(\cosh\omega x-\frac{\sinh\omega x}{\omega
x})\cong (\omega x)^2 \label{approx1}
\end{equation}
for $\omega x\ll 1$ and
\begin{equation}
\eta_1\cong\psi_2\cong\frac{3}{2}e^{\omega
x}(1-\frac{1}{\omega x})
\label{approx2}
\end{equation}
for $\omega x\gg 1$. Similarly, the approximate solution near $\sigma=K$ is obtained by
replacing $x$ in (\ref{approx1}) and (\ref{approx2}) by $y\equiv K-\sigma$.

Then, we have the WKB approximation for  $\omega\gg 1/K$
\begin{equation}
\eta_1\cong Ae^{\int^\sigma
{d\sigma^\prime\sqrt{\omega^2+e^{2\phi^\prime}(2+\delta^\prime)}}}\cong
Ae^{\omega\sigma+\frac{1}{2\omega}{\int^\sigma}d\sigma^\prime
e^{2\phi^\prime}(2+\delta^\prime)}
\end{equation}
and
\begin{equation}
\eta_2\cong Be^{\int^\sigma
{d\sigma^\prime\sqrt{\omega^2+e^{2\phi}(4+R-2\delta)}}}\cong
Ae^{\omega\sigma+\frac{1}{2\omega}{\int^\sigma}d\sigma^\prime
e^{2\phi^\prime}(4+R-2\delta)}
\end{equation}
Both WKB solutions are valid for $\omega(\sigma+K)\gg1$ and $\sigma+K<<K$, and can be matched
with (\ref{approx2}) there. We have, up to an additive constant
\begin{eqnarray}
\int^\sigma d\sigma^\prime
e^{2\phi^\prime}(2+\delta^\prime)=\int^\sigma d\sigma^\prime
e^{2\phi^\prime}(-R^\prime+R^\prime+2+\delta^\prime)\nonumber\\=\int^\sigma
d\sigma^\prime\frac{d^2\phi}{d{\sigma^\prime}^2}+\int_{-K}^\sigma
d\sigma^\prime
e^{2\phi^\prime}(R^\prime+2+\delta^\prime)=2\frac{d\phi}{d\sigma}+\int_{-K}^\sigma
d\sigma^\prime(R^\prime+2+\delta^\prime)\nonumber\\
\cong2\frac{d\phi}{d\sigma}\cong-\frac{2}{x}.
\end{eqnarray}
We find that
\begin{equation}
A=\frac{3}{2}e^{\omega K}
\end{equation}
Similarly
\begin{equation}
\int^\sigma d\sigma^\prime
e^{2\phi^\prime}(4+R-2\delta)=2\frac{d\phi}{d\sigma}+2\int_{-K}^\sigma
d\sigma^\prime e^{2\phi^\prime}(R^\prime+2+\delta^\prime)\cong2\frac{d\phi}{d\sigma}\cong-\frac{2}{x}
\end{equation}
and
\begin{equation}
B=\frac{3}{2}e^{\omega K}
\end{equation}

For $K-\sigma\ll K$ but $\omega(K-\sigma)\gg1$, we have
\begin{eqnarray}
\eta_1\cong\frac{3}{2}e^{{\omega K}+\omega(K-y)+\frac{1}{\omega
y}+\frac{1}{2\omega}\int_{-K}^{K}{d\sigma e^{2\phi}(R+2+\delta)}}
\nonumber\\\cong\frac{3}{2}e^{{2\omega
K}+\frac{1}{2\omega}\int_{-K}^{K}{d\sigma
e^{2\phi}(R+2+\delta)}}e^{-\omega y}(1+\frac{1}{\omega y})
\end{eqnarray}
where $y\equiv K-\sigma$. This is to be matched to the approximate solution
\begin{equation}
\eta_1\propto e^{-\omega y}(1+\frac{1}{\omega y})
\end{equation}
there, which is valid for $\sigma\to K^-$, i.e. $y\to 0^+$.

Define the coefficient $C_1$ such that
\begin{equation}
\eta_1\cong\frac{C_1}{y},(y\rightarrow 0^+)
\end{equation}
We have
\begin{equation}
C_1\cong\frac{3}{2}e^{{2\omega
K}+\frac{1}{2\omega}\int_{-K}^{K}{d\sigma
e^{2\phi}(R+2+\delta)}},
\end{equation}
for $\omega\gg1$.

Similarly
\begin{eqnarray}
\eta_2\cong\frac{3}{2}e^{{2\omega
K}+\frac{1}{\omega}\int_{-K}^{K}{d\sigma
e^{2\phi}(R+2-\delta)}}e^{-\omega y}(1+\frac{1}{\omega y})
\end{eqnarray}
Then
\begin{equation}
\eta_2\cong\frac{C_2}{y},(y\rightarrow 0^+)
\end{equation}
with
\begin{equation}
C_2\cong\frac{3}{2}e^{{2\omega
K}+\frac{1}{\omega}\int_{-K}^{K}{d\sigma
e^{2\phi}(R+2-\delta)}},
\end{equation}
for $\omega K\gg1$.
\subsection {Fermionic contribution}
The process is similar to the bosonic contribution part, the
equation reads
\begin{equation}
D_\pm\psi=0
\end{equation}
where
\begin{equation}
D_\pm=-\frac{d^2}{d\sigma^2}+\omega^2+e^{2\phi}\mp
e^\phi\frac{d\phi}{d\sigma}
\end{equation}
Owning to the symmetry property, we just consider
\begin{equation}
D_+\psi_+=0
\end{equation}
For $x\equiv\sigma+K\ll K$, the vanishing solution at $\sigma=-K$
reads
\begin{equation}
\eta_+\cong3(\cosh\omega x-\frac{\sinh\omega x}{\omega x})\cong
(\omega x)^2
\end{equation}
for $\omega x\ll 1$ and
\begin{equation}
\eta_+\cong\frac{3}{2}e^{\omega x}(1-\frac{1}{\omega x})
\end{equation}
for $\omega x\gg 1$

Similarly, the WKB approximation for $\omega\gg 1/K$
\begin{eqnarray}
\eta_+\cong\frac{3}{2}e^{\omega K}e^{\int^\sigma
d\sigma^\prime\sqrt{\omega^2+e^{2\phi}-e^\phi\frac{d\phi}{d\sigma}}}
\cong\frac{3}{2}e^{{\omega(K+\sigma)}+\frac{1}{2\omega}\int^\sigma
d\sigma^\prime(e^{2{\phi^\prime}}-e^{\phi^\prime}\frac{d\phi}{d\sigma^\prime})}
\nonumber\\=\frac{3}{2}e^{\omega(K+\sigma)+\frac{1}{2\omega}\frac{d\phi}{d\sigma}-\frac{1}{2\omega}e^\phi+\frac{1}{4\omega}\int_{-K}^\sigma
d\sigma^\prime e^{2{\phi^\prime}}(R^\prime+2)}
\end{eqnarray}
Define $C_+$ such that
\begin{equation}
\eta_+(K)=3C_3
\end{equation}
so we obtain that
\begin{equation}
C_3=\frac{1}{2}e^{2K\omega+\frac{1}{4\omega}\int_{-K}^K
d\sigma e^{2{\phi^\prime}}(R^\prime+2)},(\omega\gg1)
\end{equation}\\
In terms of
\begin{equation}
I\equiv\int_0^{z_0}dz\sqrt{g}(R+2)=\frac{1}{2}\int_{-K}^Kd\sigma
e^{2\phi}(R+2)
\end{equation}
\begin{equation}
J\equiv\int_0^{z_0}dz\sqrt{g}\delta=\frac{1}{2}\int_{-K}^Kd\sigma
e^{2\phi}\delta
\end{equation}
we have for $\omega\gg 1/K$
\begin{equation}
C_1(\omega)=\frac{3}{2}e^{{2K\omega}+\frac{I+J}{\omega}+...},
\end{equation}
\begin{equation}
C_2(\omega)=\frac{3}{2}e^{{2K\omega}+\frac{2(I-J)}{\omega}+...},
\end{equation}
\begin{equation}
C_3(\omega)=\frac{1}{2}e^{{2K\omega}+\frac{I}{2\omega}+...}.
\end{equation}
%


\end{document}